\documentclass[aps,twocolumn,pra,groupedaddress,superscriptaddress,nofootinbib]{revtex4-1}
\usepackage{graphicx,amsmath,amssymb,amsbsy,subfigure,hyperref,bbm,times,txfonts}

\hypersetup{
   colorlinks=true,
   linkcolor=blue,
}
\graphicspath{{figure/}}
\makeatletter
\g@addto@macro\@floatboxreset\centering
\makeatother
\begin{document}

\newcommand{\ket}[1]{|#1\rangle}
\newcommand{\bra}[1]{\langle#1|}
\newcommand{\ketbra}[1]{| #1\rangle\!\langle #1 |}
\newcommand{\kebra}[2]{| #1\rangle\!\langle #2 |}
\newcommand{\id}{\mathbbm{1}}
\newcommand{\ohm}{\Omega_{\rm CQ}}
\newcommand{\rhobd}{\rho^{\vec{c}}_{AB}}
\providecommand{\tr}[1]{\text{tr}\left[#1\right]}
\providecommand{\tra}[1]{\text{tr}_A\left[#1\right]}
\providecommand{\trb}[1]{\text{tr}_B\left[#1\right]}
\providecommand{\abs}[1]{\left|#1\right|}
\providecommand{\sprod}[2]{\langle#1|#2\rangle}
\providecommand{\expect}[2]{\bra{#2} #1 \ket{#2}}


\title{Quantum enhancement of qutrit dynamics through driving field and photonic band-gap crystal}

\author{Negar Nikdel Yousefi}
\affiliation{Department of Physics, University of Guilan, P. O. Box 41335--1914, Rasht, Iran}

\author{Ali Mortezapour}
\email{mortezapour@guilan.ac.ir }
\affiliation{Department of Physics, University of Guilan, P. O. Box 41335--1914, Rasht, Iran}

\author{Ghasem Naeimi}
\email{ghnaeimi@gmail.com}
\affiliation{Department of Physics, Qazvin Branch, Islamic Azad University, Qazvin, Iran}

\author{Farzam Nosrati}
\affiliation{Dipartimento di Ingegneria, Universit\`{a} di Palermo, Viale delle Scienze, 90128 Palermo, Italy}
\affiliation{INRS-EMT, 1650 Boulevard Lionel-Boulet, Varennes, Québec J3X 1S2, Canada}

\author{Aref Pariz}
\affiliation{Department of Biology, University of Ottawa, ON, Canada}

\author{Rosario Lo Franco}
\email{rosario.lofranco@unipa.it}
\affiliation{Dipartimento di Ingegneria, Universit\`{a} di Palermo, Viale delle Scienze, 90128 Palermo, Italy}

\begin{abstract}
A comparative study of a qutrit (three-level atomic system) coupled to a classical field in a typical Markovian reservoir (free space) and in a photonic band-gap (PBG) crystal is carried out. The aim of the study is to assess the collective impact of structured environment and classical control of the system on the dynamics of quantum coherence, non-Markovianity, and estimation of parameters which are initially encoded in the atomic state. We show that the constructive interplay of PBG material as a medium and classical driving field as a part of system results in a significant enhancement of all the quantum traits of interest, compared to the case when the driven qutrit is in a Markovian environment. Our results supply insights for preserving and enhancing quantum features in qutrit systems which are promising alternative candidates to be used in quantum processors instead of qubits.

\end{abstract}

\date{\today}


\maketitle

\section{Introduction}

In physics, no realistic quantum system is completely isolated from its surrounding environment and always there are inevitable interactions that affect the evolution of the system. As an adverse consequence of such detrimental interactions, system loses its coherence. The theory of open quantum systems deals with such systems \cite{ref1,ref2,ref3}. Since the genesis of many quantum phenomena traces back to coherence, nowadays this feature is considered as a key concept which enables tremendous possibilities in a wide spectrum of quantum technologies, quantum metrology \cite{ref4,refNaeimi,ref5,coher11}, and quantum thermodynamics \cite{ref6,ref7}. Several strategies have been then devised to protect coherence from being lost in quantum systems \cite{coher1,coher2,coher3,coher4,coher5,coher6,coher7,coher8,coher9,coher10,coher11,coher12,ref9}.

The process of losing quantum coherence, named decoherence, is usually categorized into two Markovian and non-Markovian regimes. In the Markovian regime, which is recognized as a memoryless evolution, the information leaks out to the environment irreversibly. In contrast, in non-Markovian (memory-keeping) evolution, the leaked information returns to the system \cite{ref1,ref9,ref10}. As a fundamental trait, non-Markovianity itself can be quantified by a variety of measures \cite{ref11,ref12,ref13,ref14,ref15,ref16,ref18,ref19,RivasPRL} and exploited as a resource for certain applications \cite{ref20,ref21}.

On the other hand, measurements in open quantum systems are the only way through which one can look at the quantum world to gain insight. Therefore, any advancement in quantum mechanics strongly depends on making progress in measurement techniques. The more precise measurements, the more reliable results. However, the measurement process itself can also cause decoherence and consequently reduce the accuracy of the quantum parameter estimation outcome. Increasing the degree of sensitivity and accuracy of quantum parameter estimation, exploiting quantum properties, is the primary purpose of quantum metrology \cite{ref22,refNajafi}. In recent years, this research line has been under the spotlight owing to its profound impact on quantum technology \cite{PirandolaReview}.

Quantum estimation theory provides a framework where quantum Fisher information (QFI) is employed as a reliable figure of merit to evaluate the accuracy of unknown parameters in the system. The QFI represents intrinsic information in the quantum state and is not related to the actual measurement procedure. It characterizes the maximum amount of information that can be extracted from quantum experiments about unknown parameters using ideal measurement devices \cite{Helstrom,Holevo}. Behaviors of QFI have been widely investigated both theoretically and experimentally in different systems \cite{ref28,ref29,ref30,ref31,ref32}. Decoherence always acts as a drawback, limiting the precision in the measurement outcomes \cite{ref33,ref34,ref35}. To tackle this issue, proposals to control QFI against environmental noise have been provided \cite{ref36,ref37,ref38,ref39,ref40,ref41,ref42,ref43,ref44,ref45,ref46}. 
 
 Decoherence effects may typically be weakened by engineering suitable structured environments. In this context, photonic crystals are materials possessing photonic band gap (PBG) where a range of electromagnetic frequencies are prohibited from propagating. Owing to such feature, the density of states of PBG materials substantially differ from a free space vacuum field which enables us to localize and manipulate the light within its structure\cite{ref48,Hatef}. Therefore, the mentioned difference leads to inhibition of spontaneous emission of the atoms located in PBG material\cite{ref49,ref54,ref55,ref56}. Moreover, as the atomic resonance gets close to the photonic band edge, the radiative dynamics experiences a long-time memory effect \cite{ref57,ref58,Hoeppe}. Hence, the PBG materials can be a neat solution for overcoming decoherence issue and subsequently for quantum information tasks \cite{Bellomo,Yang,Singh}.

Besides environmental engineering, classical control by driving fields can be adopted to manipulate individual quantum systems. Classical control is indeed an effective method to harness the dynamics of open quantum systems, which can be implemented in both cavity-QED and circuit-QED setups \cite{Murch,Long}.
 
To overcome limitations in controlling circuit elements for performing computational tasks via superconducting qubits, qutrits (three-level quantum systems) have been proposed as promising alternative candidates for quantum processors \cite{qutrit1,qutrit2}. It is noteworthy that multilevel systems reduce the number of required circuit elements through extending the Hilbert space \cite{qutrit14}. This characteristic offers interesting possibilities for novel fundamental tests of quantum mechanics \cite{qutrit3,qutrit4}, increased security in a range of quantum information protocols \cite{qutrit5,qutrit6,qutrit7,qutrit8,qutrit9,qutrit10,qutrit11}, larger channel capacity in quantum communication \cite{qutrit12}, and more efficient quantum gates \cite{qutrit13,qutrit14}. 
It is thus of particular interest to increase our knowledge about the dynamical behavior of the quantum features of a qutrit under suitable environmental conditions.  
  
In this work, we consider a classically driven three-level atomic system as a qutrit which is placed in either free space or a PBG crystal (structured reservoir). This way, we can make a comparative study and individuate the conditions for the enhancement of the quantum properties of interest. In particular, we assess the influence of both driving laser field and PBG reservoir on the time evolution of quantum coherence, non-Markovianity and QFI of the qutrit. Such a comprehensive investigation supplies useful insights about the possibility of preserving and controlling quantumness in a three-level open quantum system which can be employed as a constituent of a qutrit-based register.     

The paper is organized as follows: In Sec.~\ref{secMS}, we describe the model and give explicit expression to the evolved reduced density matrix of the atomic system for two considered situations; the atom in free space and the atom in photonic band gap. In Sec.~\ref{sec:QC}, the time evolution of coherence for both situations are comparatively discussed. Sec.~\ref{sec:NM} presents the results concerning the non-Markovian dynamics of the system by employing the HSS measure. The dynamics of quantum Fisher information and optimal parameters estimation is reported in Sec.~\ref{sec:QFI}. In Sec.~\ref{sec:Disc} we discuss the experimental context where our results can be reproduced. Conclusive remarks and perspectives of this work are summarized in Sec.~\ref{sec:Conc}.

\section{Model and Solution}\label{secMS}

\begin{figure}[t!]
\includegraphics[width=0.3\textwidth]{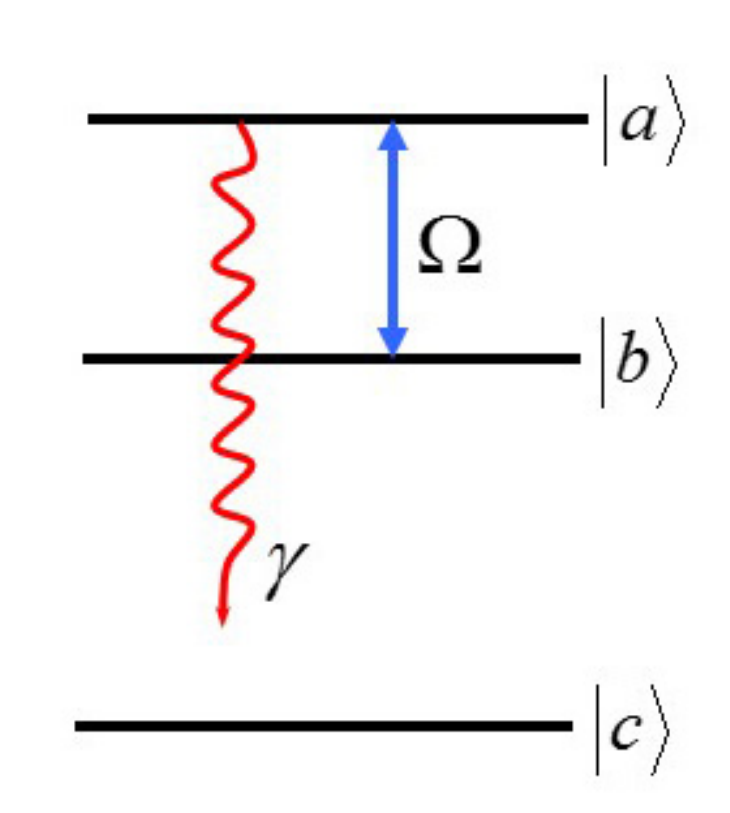}
\caption{Illustration of the energy levels of the atomic qutrit. Atomic states of the system are characterized by $\left| a \right\rangle$, $\left| b\right\rangle$ and $\left| c \right\rangle$. The double arrow and the wavy arrow respectively denote the coupling laser field with Rabi frequency $\Omega$ and the spontaneous emission of rate $\gamma$.}
\label{fig1}
\end{figure}

We consider a three-level atom which interacts with the vacuum field and a driving field. As schematically depicted in Fig.~\ref{fig1}, the upper level $\left| a \right\rangle$ decays to the ground level $\left| c \right\rangle$ with rate $\gamma$ due to the interaction with vacuum reservoir modes. Meanwhile, the transition $\left| a \right\rangle \leftrightarrow \left| b \right\rangle$ is resonantly coupled by means of a coherent laser field with the Rabi frequency $\Omega$. Such a system is recognized as upper-level coupling \cite{Scully,Mortezapour1}. The Hamiltonian of the system in the interaction picture can be written as ($\hbar\equiv1$) 
\begin{equation}
\label{eq:1}
\hat{V}=\Omega \left| a \right\rangle \left\langle b \right|+ \sum_{k}g_{k} \left| a \right\rangle \left\langle c \right| \hat{b}_k e^{i(\omega_{ac}-\omega_k)t}+c.c,
\end{equation}
where $\omega_{ab}$, $\omega_{ac}$ are the frequencies of $\left| a \right\rangle \rightarrow \left| b \right\rangle$ and $\left| a \right\rangle \rightarrow \left| c \right\rangle$ transitions, respectively; $b_k(b_{k}^{\dagger})$ indicates annihilation (creation) operator for the $k$-th vacuum mode with frequency $\omega_k$. The parameter $\Omega$ characterizes the Rabi frequency of the coupling laser field. Here $g_k$ (assumed as a real number) denotes the coupling constant between the $k$-th vacuum mode and the atomic transition $\left| a \right\rangle \rightarrow \left| c \right\rangle$. Its expression is
\begin{equation}
\label{eq:new}
g_{k}=\dfrac{\omega_{ac}d_{ac}}{\hbar}  \left(\frac{\hbar}{2\epsilon_{0}\omega_{k}V}\right)^{1/2} \hat{e}_{k}\cdot\hat{d}_{ac},
\end{equation}
where $d_{ac}$ and $\hat{d_{ac}}$ are the magnitude and unit vector of the
atomic dipole moment for the transition $\left| a \right\rangle \rightarrow \left| c \right\rangle$, $V$ is the sample volume, $\hat{e}_{k}$ is the transverse polarization unit vector, and $\epsilon_{0}$ is the Coulomb constant.

We assume the atom to be initially in a pure superposition of $\left| a \right\rangle$ and $\left| b \right\rangle$,
\begin{equation}
\label{eq:2}
\left| \Psi_\mathrm{A}(0) \right\rangle = \cos(\theta/2)\left| a \right\rangle + e^{i\phi} \sin(\theta/2)\left| b \right\rangle,
\end{equation}
and the field in the vacuum state $\left| 0 \right\rangle$. Hence, at any later time $t$, the quantum state of the whole system can be described as
\begin{equation}
\label{eq:3}
\left| \Psi_\mathrm{AF}(t) \right\rangle =A(t)\left| a \right\rangle \left| 0 \right\rangle+B(t)\left| b \right\rangle \left| 0 \right\rangle+  \sum_{k}C_k(t)\left| c \right\rangle \left| 1_k \right\rangle,
\end{equation}
where $\left| 1_k \right\rangle$ indicates the state with one photon in $k$-th vacuum mode. Substituting Eq.~\eqref{eq:3} into the Schr\"{o}dinger equation, we obtain the following set of coupled equations for the probability amplitudes $A(t)$, $B(t)$, and $C_k(t)$
\begin{subequations}
\begin{align}
\label{eq:4}
\dot{A}(t)&=-i\Omega^*B(t)-i\sum_{k}g_kC_k(t)e^{-i\delta_kt},\\
\label{eq:5}
\dot{B}(t)&=-i\Omega A(t),\\
\label{eq:6}
\dot{C}(t)&=-ig_kA(t)e^{i\delta_kt},
\end{align}
\end{subequations}
where $\delta_k=\omega_k-\omega_{ac}$ is the detuning of the radiation mode frequency $\omega_k$ from the atomic transition frequency $\omega_{ac}$. Solving Eq.~\eqref{eq:6} formally and substituting the solution into Eq.~\eqref{eq:4}, one obtains 
\begin{equation}
\label{eq:7}
\dot{A}(t)=-i\Omega B(t)-\int^{t}_{0}F(t,t')A(t')dt'
\end{equation}
where
\begin{equation}
\label{eq:8}
F(t-t')=\sum_{k}g_k^2e^{-i\delta_k(t-t')}.
\end{equation}
is the correlation function including the memory effects induced by the reservoir. The memory kernel strictly depends on the spectral density of the field in the reservoir.

The evolved reduced density matrix $\rho(t)$ of the driven atomic qutrit is straightforwardly obtained by tracing out the environmental degrees of freedom in Eq.~\eqref{eq:3}. In the basis $\{ \left| a \right\rangle, \left| b \right\rangle, \left| c \right\rangle \}$, it is given by
\begin{equation}
\label{eq:24}
\rho(t)=
\begin{pmatrix}
\rho_{aa} & \rho_{ab} & \rho_{ac}\\
\rho_{ba} & \rho_{bb} & \rho_{bc}\\
\rho_{ca} & \rho_{cb} & \rho_{cc}
\end{pmatrix},
\end{equation}
with
\begin{align}
\label{eq:25}
\begin{split}
\rho_{aa} &= \left | A(t) \right | ^2, \
\rho_{bb} = \left | B(t) \right | ^2, \
\rho_{cc} = 1-\left | A(t) \right | ^2-\left | B(t) \right | ^2, \\
\rho_{ab} &=\rho^{*}_{ba}=A(t)B^*(t),\
\rho_{ac} = \rho^*_{ca}=0,\
\rho_{bc} = \rho^*_{cb}=0,
\end{split}
\end{align}

In the following, we give the solutions for the time-dependent amplitudes in the two cases of interest: free space and photonic crystal.

\subsection{Qutrit in free space}
Let us assume the atom is located in free space, i.e., a broadband reservoir with the photon dispersion relation $\omega_k=ck$. One can thus use the Weisskopf-Wigner approximation \cite{Weisskopf} to obtain $F(t-t')=\frac{\gamma}{2}\delta(t-t')$ (no memory effects), with $\gamma=\frac{1}{4\pi\varepsilon_0}(\frac{4\omega_{ac}^3 \left | d_{ac}\right | ^2}{6\hbar c^3})$ being the spontaneous emission rate from level $\left| a \right\rangle$ to level $\left| c \right\rangle$. Simultaneously solving Eqs.~\eqref{eq:5}, \eqref{eq:6} and \eqref{eq:8} yields the amplitudes
\begin{subequations}
\label{eq:9}
\begin{align}
A(t)&=A_1e^{y_1t}+A_2e^{y_2t},\\
B(t)&=B_1e^{y_1t}+B_2e^{y_2t},\\
C_k(t)&=-ig_k\left[A_1\frac{e^{(y_1+i\delta_k)t}-1}{y_1+i\delta_k}+A_2\frac{e^{(y_2+i\delta_k)t}-1}{y_2+i\delta_k}\right],
\end{align}
\end{subequations}
where, defining $\beta=[(\gamma/2)^2-4\left|\Omega\right| ^2]^{1/2}$, 
\begin{align}
\label{eq:10}
\begin{split}
y_{1,2} &=-[(\gamma/2)\pm\beta]/2, \\
A_1 &=-[y_1\cos(\theta/2)-i\Omega e^{i\phi}\sin(\theta/2)]/\beta,\\
A_2 &=\cos(\theta/2)-A_1,\\
B_1 &=-[\sin(\theta/2)(y_1+\gamma/2)e^{i\phi}-i\Omega^*\cos(\theta/2)]/\beta,\\
B_2 &=e^{i\phi}\sin(\theta/2)-B_1.
\end{split}
\end{align}
Putting the above solutions for $A(t)$ and $B(t)$ in Eq.~\eqref{eq:25}, we get the evolved reduced density matrix $\rho(t)$ of the driven qutrit in the free space.

\subsection{Qutrit in a photonic crystal}
We now divert our attention to the case in which the three-level atom is embedded in a 3D PBG material, assuming the transition frequency $\omega_{ac}$ is near the edge of a photonic band gap \cite{ref54,ref55,ref75,ref76,ref77,ref78}. Regarding this situation, and owing to the rapid change of the density of electromagnetic modes in the vicinity of the atomic transition frequency, the Weisskopf-Wigner approximation is no longer valid. Therefore, a more rigorous relation is required instead of Eq.~\eqref{eq:8}. It is well-known that, in a real 3D PBG material with an allowed point-group symmetry, the gap is highly anisotropic and the photon dispersion relation in the effective-mass approximation gets the form \cite{ref77,ref78}
\begin{equation}
\label{eq:11}
\omega_{\vec{k}}=\omega_c+A(\vec{k}-\vec{k}_0)^2, \mspace{18mu} A\approx fc^2/\omega_c^2,
\end{equation}
where $\omega_c$ is the upper band edge frequency and $\vec{k}$ denotes the wavevector; $\vec{k}_0$ is a specific wavevector related to the point-group symmetry of the dielectric material with modulus $k_0\equiv\pi/L$ with $L$ being the lattice constant of the photonic crystal. Also, $f$ is a dimensionless scaling factor, whose value depends on the nature of the dispersion relation near the band edge $\omega_c$. The anisotropic effective mass dispersion (Eq.~\eqref{eq:11}) leads to a photonic density of states at a band edge $\omega_c$ which behaves as $J(\omega)\approx (\omega-\omega_c)^{1/2}$, for $\omega >\omega_c$, characteristic of a 3D phase space \cite{ref75}. This dispersion relation is valid for frequencies close to the upper photonic band edge. 

Using the anisotropic effective-mass dispersion relation Eq.~\eqref{eq:11} and assuming  and assuming $(t-t')$ is large enough to satisfy the condition $\omega_c(t-t')\gg 1$, the kernel in the continuum limit can be derived as \cite{ref77,ref78}
\begin{equation}
\label{eq:12}
F(t-t')=-\alpha\frac{e^{i[\delta(t-t')+\pi/4]}}{\sqrt{4\pi(t-t')^3}}, \mspace{18mu} \omega_c(t-t') \gg1,
\end{equation}
where
\begin{equation}
\label{eq:13}
\alpha^2\approx\frac{\omega_c}{16f^3}\left(\frac{\gamma}{\omega_{ac}}\right)^2,
\end{equation}
has the dimension of a frequency and $\delta=\omega_{ac}-\omega_c$ denotes the detuning of the atomic transition frequency $\omega_{ac}$ from the upper band edge frequency $\omega_c$. In contrast to the free space case, Eq.~\eqref{eq:12} explicitly keeps memory of the past history (times) of the system. Hence, it describes memory effects in the spontaneous emission dynamics due to the presence of the photonic band gap. In other words, the atom-reservoir interaction within a PBG is expected to be non-Markovian.

By taking the Laplace transforms of Eqs.~\eqref{eq:7} and \eqref{eq:5} and using the initial state of Eq.~\eqref{eq:2}, we obtain  
\begin{equation}
\label{eq:14}
\tilde{A}(s+i\delta)=\frac{(s+i\delta)\cos(\theta/2)-\Omega e^{i\phi}\sin(\theta/2)}{D(s)},
\end{equation}
\begin{equation}
\label{eq:15}
\tilde{B}(s+i\delta)=[(s+\alpha e^{i\pi/4}\sqrt{s} +i\delta)e^{i\phi}\sin(\theta/2)+\Omega \cos(\theta/2)]/D(s),
\end{equation}
with $D(s)=(s+i\delta)^2+\alpha e^{i\pi/4}(s+i\delta)\sqrt{s}+\Omega^2=\displaystyle\prod_{j=1}^{4}(\sqrt{s}-e^{i\pi/4}u_j)$. Here, $u_j$ $(j=1,...,4)$ are the roots of the quartic equation
\begin{equation}
\label{eq:16}
x^4+\alpha x^3+2\delta x^2+\alpha \delta x-(\Omega^2-\delta^2)=0,
\end{equation}
which are given by
\begin{align}
\label{eq:17}
\begin{split}
u_{1,3}&=-\sigma_1 \pm \sqrt{E-r/2+\sigma^2_1}, \\
u_{2}&=u^*_4=-\sigma_2 -i\sqrt{E+r/2-\sigma_2^2},
\end{split}
\end{align}
where
\begin{subequations}
\label{eq:18}
\begin{align}
\sigma_{1,2}&=(\alpha\pm \sqrt{\alpha^2-8\delta+4r})/4,\\
E&=(r^2/4+\Omega^2-\delta^2)^{1/2},\\
r&=(M-q/2)^{1/3}-(M+q/2)^{1/3}+\eta_1/3,
\end{align}
\end{subequations}
with
\begin{align}
\label{eq:19}
\begin{split}
M&=\left[\left(\frac{P}{3}\right)^3+\left(\frac{q}{2}\right)^2\right]^{1/2}, \
P=-\frac{\eta^2_1}{3}+\eta_2,\\
q&=-2\left(\frac{\eta_1}{3}\right)^3+\frac{\eta_1\eta_2}{3}+\eta_3,
\end{split}
\end{align}
and
\begin{align}
\label{eq:20}
\begin{split}
\eta_1&=2\delta, \\
\eta_2&=\alpha^2\delta+4(\Omega^2-\delta^2),\\
\eta_3&=(\alpha^2-8\delta)(\Omega^2-\delta^2)-\alpha^2\delta^2.
\end{split}
\end{align}
Numerical analysis shows that the roots $u_{1,3}$ are real ($u_1$ is positive but $u_2$ is negative), and the roots $u_{2,4}$ are complex conjugates of each other with a negative real part ($u_2$ and $u_4$ lie in the third and second quadrants, respectively). The probability amplitudes $A(t)$ and $B(t)$ are determined by inverting Eqs.~\eqref{eq:14} and \eqref{eq:15} via the complex inversion formula, that is
\begin{equation}
\label{eq:21}
A(t)=\sum_{j=1}^{2}p_jQ_{3j}e^{i(u_j^2+\delta)t}+\frac{\alpha e^{i\pi/4}}{\pi}\int^{\infty}_{0}\frac{g_3(x)e^{-(x-i\delta)t}}{Z(X)}dx,
\end{equation}
\begin{equation}
\label{eq:22}
B(t)=\sum_{j=1}^{2}p_jQ_{2j}e^{i(u_j^2+\delta)t}+\frac{\alpha\Omega e^{i\pi/4}}{\pi}\int^{\infty}_{0}\frac{g_2(x)e^{-(x-i\delta)t}}{Z(X)}dx,
\end{equation}
where
\begin{subequations}
\label{eq:23}
\begin{align}
p_j&=\frac{2u_j}{(u_j-u_l)(u_j-u_m)(u_j-u_n)},\\
& (l,m,n=1,...,4 j\neq1\neq m\neq n),\nonumber\\
Q_{3j}&=(u_j^2+\delta)\cos(\theta/2)+i\Omega e ^{i\phi}\sin(\theta/2),\\
Q_{2j}&=(u_j^2+\alpha u_j+\delta)e^{i\phi}\sin(\theta/2)-i\Omega \cos(\theta/2),\\
g_3(x)&=[(-x+i\delta)\cos(\theta/2)-\Omega e^{i\phi} \sin(\theta/2)](-x+i\delta)\sqrt{x},\\
g_2(x)&=[(-x+i\delta)\cos(\theta/2)-\Omega e^{i\phi}\sin(\theta/2)]\sqrt{x},\\
Z(x)&=[(-x+i\delta)^2+\Omega ^2]^2+i\alpha^2 (-x+i\delta)^2x.
\end{align}
\end{subequations}
Having $A(t)$ and $B(t)$, from Eq.~\eqref{eq:25} we obtain the evolved reduced density matrix $\rho(t)$ of the driven qutrit in the photonic crystal. 


Knowledge of the evolved reduced density matrix of the atomic qutrit shall allow us to study the dynamics of the quantum properties of interest, comparing them to the case of free space. This analysis is reported in the following sections.

\section{Quantum Coherence}\label{sec:QC}

Quantum coherence represents the coherent superposition of distinct physical states which draw fundamental distinction between quantum mechanics and classical physics. It is also a resource for quantum information processing \cite{coher11}. Hence, it is important investigate the dynamical behavior of quantum coherence in basic systems which can be promising constituents of quantum registers, such as the qutrit system described above. In this section we perform such a study.  

\begin{figure}[t!]
\centering
\includegraphics[width=0.52\textwidth]{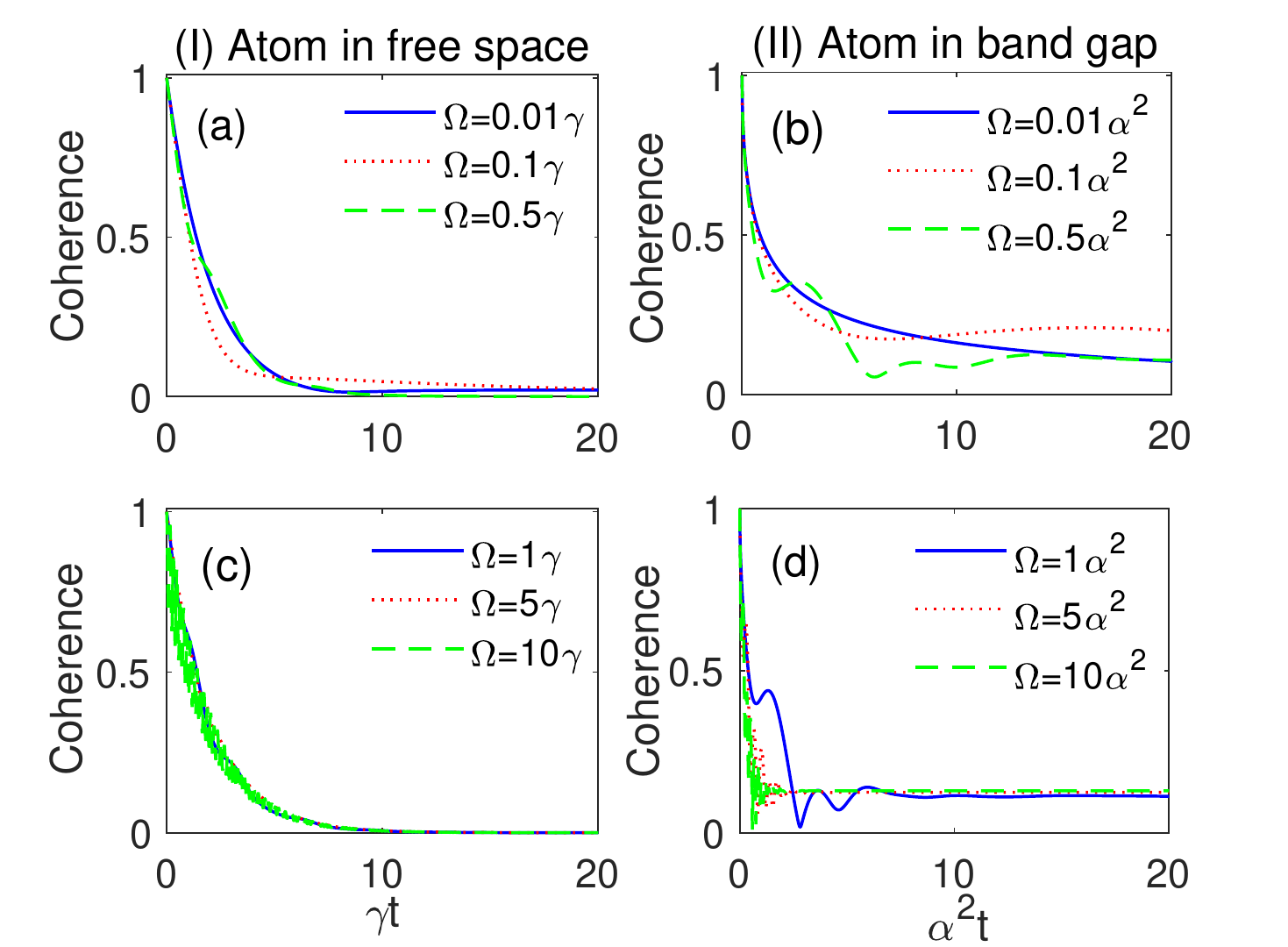}
\caption{Dynamics of qutrit coherence for different Rabi frequencies $\Omega$ of the driving laser field. Column (I) and column (II) correspond to the case when the atom is placed, respectively, in free space and in a band-gap material with $\delta=0$. Values of the initial state parameters are $\theta=\pi/2$, $\phi=\pi/4$.}
\label{fig2}
\end{figure}

To quantify coherence we employ the $l_1$ norm of coherence, which is defined as a sum of the absolute values of all off-diagonal elements in the density matrix $\rho_{ij}$ using following expression \cite{ref79}
\begin{equation}
\label{eq:28}
C_{l_1}(\rho(t))=\sum_{i\neq j}\left|\rho_{ij}(t)\right|,
\end{equation}
where $|\rho_{ij}(t)|$ are the absolute values of all off-diagonal elements of the qutrit density matrix $\rho(t)$ of Eq.~\eqref{eq:24}. Notice that the coherence of the initial pure state $\left| \Psi_\mathrm{A}(0) \right\rangle$ of Eq.~\eqref{eq:2} is $C_{l_1}(\rho(0))=\sin\theta$. The angle $\theta$ of the initial state is fixed to $\theta=\pi/2$ to maximize the initial quantum coherence of the qutrit, $C_{l_1}(\rho(0))=1$.

Fig.~\ref{fig2} shows the quantum coherence as a function of the scaled time for different Rabi frequencies of the driving laser field. The two situations are compared in which the atom is placed in either free space or photonic band gap; the corresponding plots are presented, respectively, in column (I) and column (II) of the figure. When the atom is located in a free space, it is seen that the coherence of the system rapidly diminishes and vanishes in a short while. This behavior is typical for all the intensities of the laser field, since it stems from the leakage of quantum information from the quantum system to the environment. Contrarily, for the case in which the atom is placed in a photonic band-gap material, the qutrit loses coherence to some extent and reaches a non-zero steady-state value. Furthermore, it is clear for both cases that as $\Omega$ increases, quantum coherence dynamics manifests a nonmonotonic behavior with oscillations. 

\begin{figure}[t!]
\centering
\includegraphics[width=0.52\textwidth]{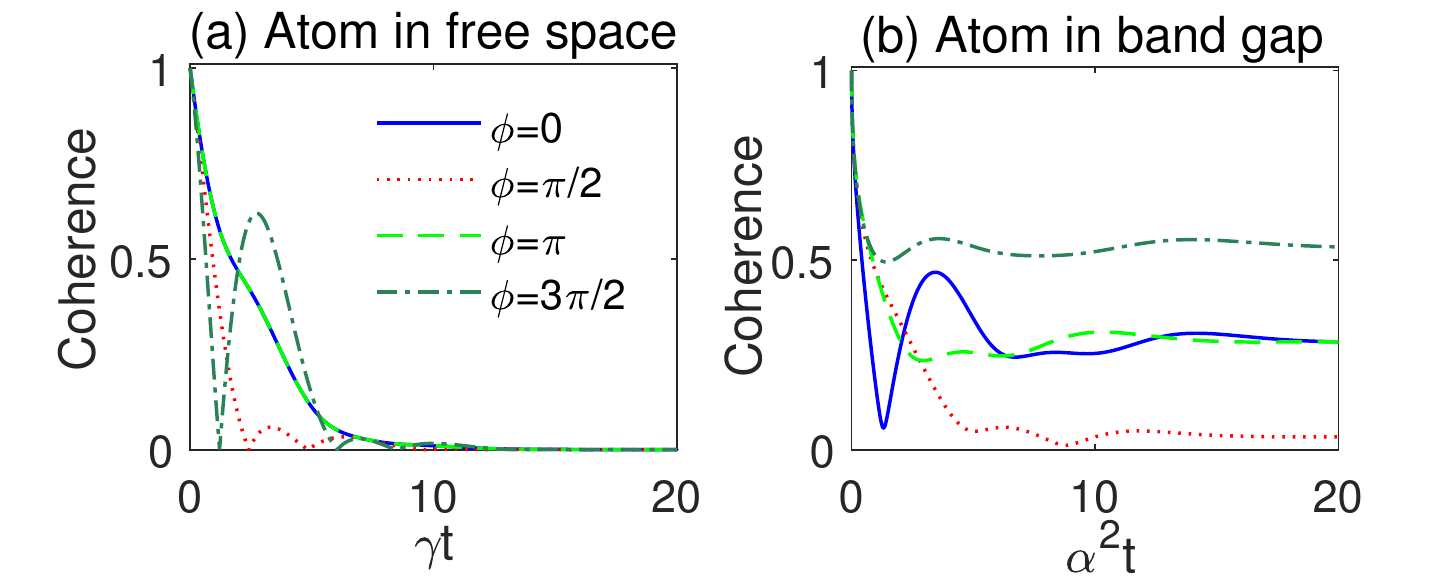}
\caption{Dynamics of qutrit coherence for different values of $\phi$ with $\theta=\pi/2$. (a) Atom in free space with $\Omega=0.5\gamma$; (b) atom in band-gap material with $\Omega=0.5\alpha^2$, $\delta=0$.}
\label{fig3}
\end{figure}

Fig.~\ref{fig3} displays the time evolution of qutrit coherence for various values of the initial relative phase $\phi$. This analysis is useful to supply the dependence of the dynamics on the relative phase assigned to the initial state of the qutrit. As is evident, the initial relative phase significantly affects the dynamic behavior of quantum coherence in both the environmental conditions. Although the dynamics exhibits the same pattern at the beginning in both media, it radically changes at longer times. The initial fluctuating dynamics depends on $\phi$ for both cases, but quantum coherence in free space eventually vanishes regardless of the initial relative phase. Instead, in the case of the photonic band-gap crystal, quantum coherence tends to a steady-state value whose amount is ruled by the value of $\phi$. 

The main message of this analysis is clear: while the photonic crystal as a structured reservoir can guarantee a stationary quantum coherence of the driven qutrit, the relative phase of the initial state has to be opportunely chosen in order to maintain it to a desired extent. Instead, the Rabi frequency of the driving laser does not seem to have significant effects to the long-time behavior of coherence.

\section{Non-Markovianity}\label{sec:NM}

The appearance of oscillating behavior in the dynamics of quantum coherence, shown in Figs.~\ref{fig2} and \ref{fig3}, motivates us to find out whether these oscillations are due to non-Markovian features or not. This is the objective of this section.

We quantify the non-Markovian dynamics of the system by means of the Hilbert-Schmidt speed measure (HSS), which has been introduced recently \cite{ref18}. In the following, we briefly recall the gist of this non-Markovianity measure. Introducing the distance measure \cite{ref81}
\begin{equation}
\label{eq:29}
\left[d(p,q)\right ]^2=\frac{1}{2}\sum_{x}\left| p_x,q_x \right|^2,
\end{equation}
where $p=\{p_x\}_x$ as well as $q=\{q_x\}_x$ are probability distributions, and subsequently considering the classical statistical speed
\begin{equation}
\label{eq:30}
s\left[p(\phi_0)\right ]=\frac{d}{d\phi}d\left(p(\phi_0+\phi),p(\phi_0)\right),
\end{equation}
one can define a special kind of quantum statistical speed called HSS by extending these classical notions to the quantum case.
We assume an arbitrary pair of quantum states $\rho$ and $\sigma$ from the positive-operator-valued measure (POVM) which respectively possess the measurement probabilities $P_x=\mathrm{Tr}\{E_x\rho\}$ and $q_x=\mathrm{Tr}\{E_x\sigma\}$. Let us notify that the POVM is defined by the $\{E_x\ge0\}$ and meets $\sum_xE_x=I$ condition. The maximization of the classical distance of Eq.~\eqref{eq:29} over all possible choices of POVMs yields the associated quantum distance called  Hilbert-Schmidt distance \cite{ref82}
\begin{equation}
\label{eq:31}
D(\rho, \sigma)\equiv \max\limits_{E_x}d(\rho,\sigma)=\sqrt{\frac{1}{2}\mathrm{Tr}[(\rho,\sigma)^2]}.
\end{equation}
Likewise, the corresponding quantum statistical speed (HSS) can be obtained by maximizing the classical statistical speed of Eq.~\eqref{eq:30} over all possible POVMs \cite{ref18}
\begin{equation}
\label{eq:32}
HSS(\rho(\phi))\equiv \max\limits_{E_x}s[P(\phi)]\\
=\sqrt{\frac{1}{2}\mathrm{Tr}\left[\left(\frac{d\rho(\phi)}{d\phi}\right)^2\right]}.
\end{equation}
Thus, HSS can be conveniently determined using this expression in which there is no need to diagonalize $d\rho(\phi)/d\phi$. 

Given a quantum system with a $n$-dimensional Hilbert space and initial state 
\begin{equation}
\label{eq:33}
\left| \Psi_{0} \right\rangle=\frac{1}{\sqrt{n}}\left(e^{i\phi}\left| \psi_{1} \right\rangle+......+\left| \psi_{n} \right\rangle\right),
\end{equation}
where $\phi$ is a relative phase and $\left\{\left| \psi_{i} \right\rangle, i=1,2,....,n\right\}$ form a complete and orthonormal basis, the time derivative of HSS $\chi(t)=dHSS(\rho_{\phi}(t))/dt$ can be interpreted as a bona-fide witness of information flow between the system and its environment \cite{ref18}. In particular, $\chi(t)\geq0$ stands for an irreversible flow of information from the system to the environment, identifying a Markovian regime. In contrast, $\chi(t)<0$ indicates a backflow of quantum information from the environment to the system which identifies a non-Markovian regime.

To suitably investigate non-Markovianity by the HSS dynamics for the considered qutrit, the latter has to be initially set in a state of the form of Eq.~\eqref{eq:33}, that is
\begin{equation}
\label{eq:new2}
\left| \Psi_\mathrm{A}(0) \right\rangle = \frac{1}{\sqrt{3}}( \left| a \right\rangle + e^{i\phi} \left| b \right\rangle + \left| c \right\rangle ).
\end{equation}
Such an initial state leads to a density matrix $\rho(t)$ of the form of Eq.~\eqref{eq:24}, whose elements are given by Eq.~\eqref{eq:25} with the difference that now
\begin{align}
\label{eq:new3}
\begin{split}
\rho_{ac} = \rho^*_{ca}=\frac{A(t)}{\sqrt{3}},\
\rho_{bc} = \rho^*_{cb}=\frac{B(t)}{\sqrt{3}},
\end{split}
\end{align}
and $A(t)$, $B(t)$ are obtained from Eqs.~\eqref{eq:9}-\eqref{eq:10} (atom in free space) and from Eqs.~\eqref{eq:14}-\eqref{eq:23} (atom in photonic crystal) by substituting $\cos(\theta/2)=\sin(\theta/2)\equiv 1/\sqrt{3}$.

\begin{figure}[t!]
\centering
\includegraphics[width=0.52\textwidth]{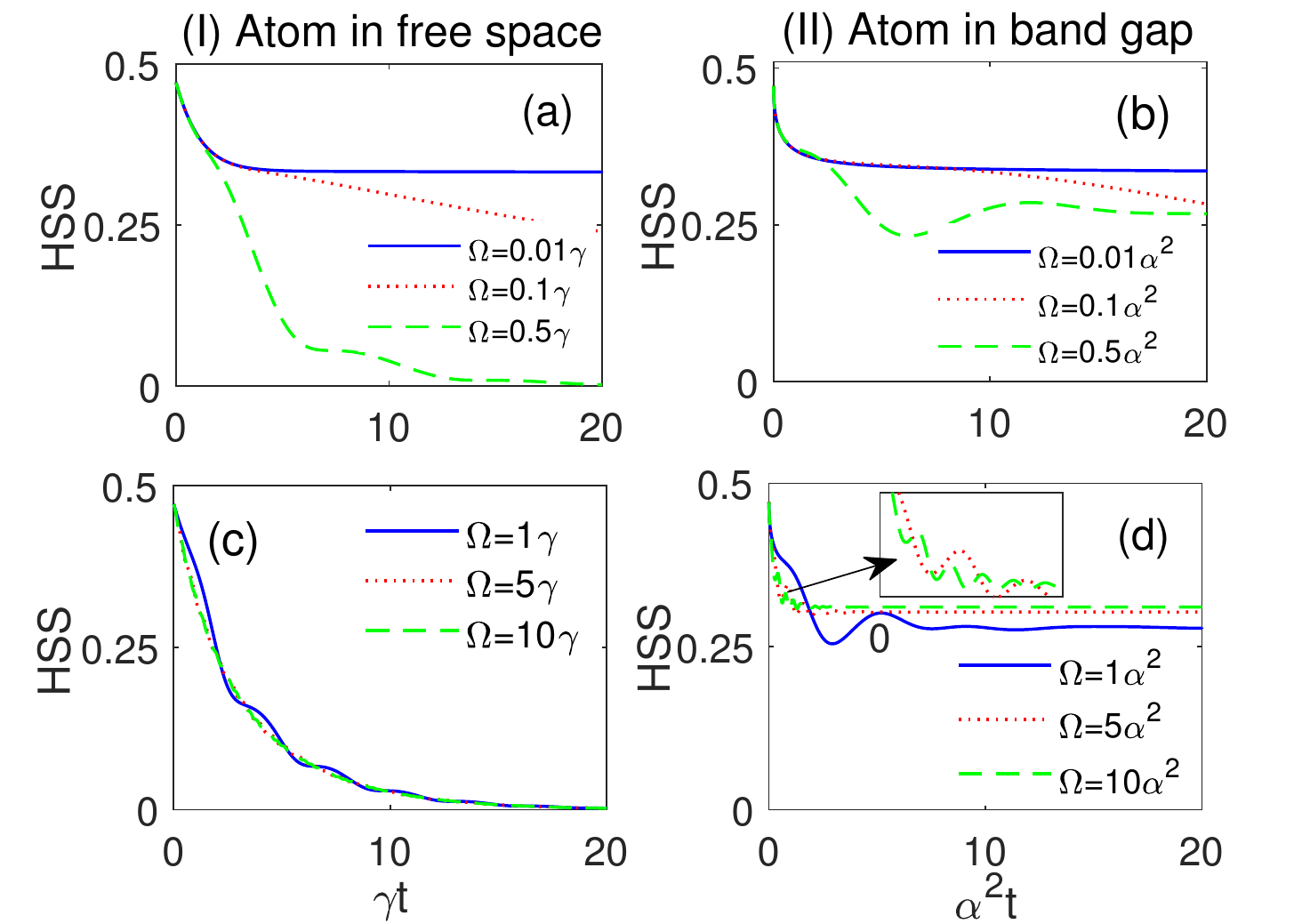}
\caption{Dynamical behavior of HSS for different Rabi frequencies $\Omega$ of the coupling laser field. Column (I) corresponds to the atom placed in free space and column (II) to the atom in a photonic band-gap material with $\delta =0$.}
\label{fig4}
\end{figure}

Fig.~\ref{fig4} displays the HSS dynamics for different Rabi frequencies $\Omega$ of the coupling laser field. It is seen that for the case of free space, the qutrit evolution exhibits a Markovian behavior regardless of the value of $\Omega$. In fact, the HSS always has a monotonic behavior with no change of its time derivative $\chi(t)$. On the other hand, the HSS dynamics in the case of photonic band-gap material quickly reacts to the increase of the intensity ($\Omega$) of the driving field (nonmonotonic curves). Moreover, a larger Rabi frequency of the coupling laser field not only gives rise to non-Markovian behavior, but it also affects the time duration of non-Markovianity.

\begin{figure}[t!]
\centering
\includegraphics[width=0.52\textwidth]{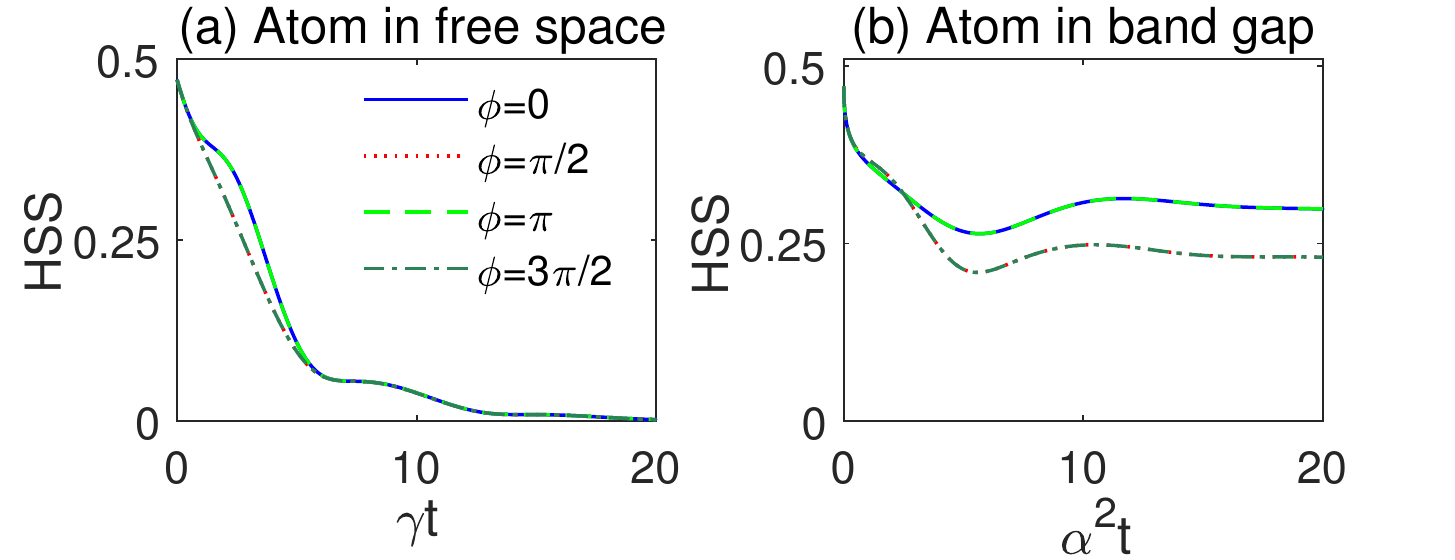}
\caption{Dynamics of HSS for different values of $\phi$. (a) Atom in free space with $\Omega=0.5\gamma$; (b) atom in band-gap material with $\Omega=0.5\alpha^2$, $\delta=0$.}
\label{fig5}
\end{figure}

To finalize the investigation about non-Markovianity, we now assess the impact of the initial relative phase $\phi$ of the state vector on the time evolution of the HSS. This is shown in Fig.~\ref{fig5}. It is observed that both situations (free space and photonic crystal) exhibit the same dynamics of the HSS for the pairs of phases $\phi=0 ,\pi$ and $\phi=\pi/2 , 3\pi/2$. As expected, when the qutrit is placed in free space, phase change cannot induce a non-Markovian behavior anywise. On the other hand, for the situated qutrit in photonic crystal, phase change has a minor effect on the non-Markovian behaviour of system. Thus, as a general characteristic, one deduces that the relative phase $\phi$ does not significantly affect the memory effects of the system dynamics neither in free space nor in PBG.

Our analysis clearly demonstrate that the photonic crystal as a structured reservoir enhances the non-Markovianity of the system dynamics. The study performed in this section provides the general tools and some values of the system parameters to quantitatively manipulate the emergence of dynamical memory effects.

\section{Quantum Fisher Information}\label{sec:QFI}

In this section, we enter the context of quantum metrology, being interested to dynamical variations in the estimation precision of the angle parameters $\phi$ and $\theta$ encoded into the initial state of the atomic qutrit defined in Eq.~\eqref{eq:2}. In particular, we aim at figuring out how changes of both Rabi frequency $\Omega$ of the coupling laser field and initial values of the relative phase $\phi$ affect the sensitivity in the measurement of $\phi$, $\theta$. 
Multiparameter quantum estimation theory allows us to deal with such a study \cite{ref83}. 

Based on this theory, the precision of simultaneous estimation of the two unknown parameters $\theta$ and $\phi$ is limited by the quantum Cramer-Rao bound (QCRB) as \cite{Helstrom,Holevo}
\begin{equation}
\label{eq:34}
{\sum}\geq {\sum}_\mathrm{min} = F^{-1}(\theta, \phi),
\end{equation}
where $\sum$ is the covariance matrix for the parameters $\theta$ and $\phi$, $F^{-1}(\theta, \phi)$ is the inverse matrix of the quantum Fisher information matrix (QFIM) $F(\theta, \phi)$. The latter is given by 
\begin{equation}
\label{eq:35}
F(\theta, \phi)=
\begin{pmatrix}
F_{\theta}(t) & F_{\theta\phi}(t) \\
F_{\phi\theta} & F_{\phi(t)}
\end{pmatrix},
\end{equation}
with $F_{\theta}=\mathrm{Tr}[\rho(t)L^2_{\theta}]$, $F_{\phi}=\mathrm{Tr}[\rho(t)L^2_{\phi}]$ and $F_{\theta\phi}(t)=F_{\phi\theta}=\frac{1}{2}\mathrm{Tr}[\rho(t)(L_{\theta}L_{\phi}+L_{\phi}L_{\theta})]$, where $L_{\theta}$ and $L_{\phi}$ are the symmetric logarithmic derivatives for the parameters $\theta$ and $\phi$ defined by
\begin{align}
\label{eq:36}
\begin{split}
\frac{\partial}{\partial \theta}\rho(t)=\frac{1}{2}\left [L_{\theta}\rho(t)+\rho(t)L_{\theta}\right ],\\
\frac{\partial}{\partial \phi}\rho(t)=\frac{1}{2}\left [L_{\phi}\rho(t)+\rho(t)L_{\phi}\right ],
\end{split}
\end{align}
respectively \cite{ref84}. Since $L_{\theta}=L_{\theta}^{\dagger}$ and $L_{\phi}=L_{\phi}^{\dagger}$, the QFIM $F(\theta, \phi)$ is Hermitian. Notice that a key feature of this matrix is to impose a lower bound to the mean-square error of any unbiased estimator for the parameters through the Cramer-Rao inequality \cite{Helstrom,Holevo}.

\subsection{Single-parameter Fisher information}
We first analyze the time evolution of the individual QFIs $F_{\phi}$ and $F_{\theta}$ related to the parameters $\phi$ and $\theta$, respectively. It is noteworthy that in the single-parameter quantum estimation, the corresponding quantum Cramer-Rao bounds for independent estimations of the parameters $\phi$ and $\theta$ are \cite{Helstrom,Holevo}
\begin{equation}
\label{eq:37}
\delta \phi \geq 1/\sqrt{F_{\phi}},\quad
\delta \theta\geq 1/\sqrt{F_{\theta}}.
\end{equation}
Therefore, to have a better measurement precision, we look for the conditions that maintain the QFIs as high as possible during the time evolution.

\begin{figure}[t!]
\centering
\includegraphics[width=0.52\textwidth]{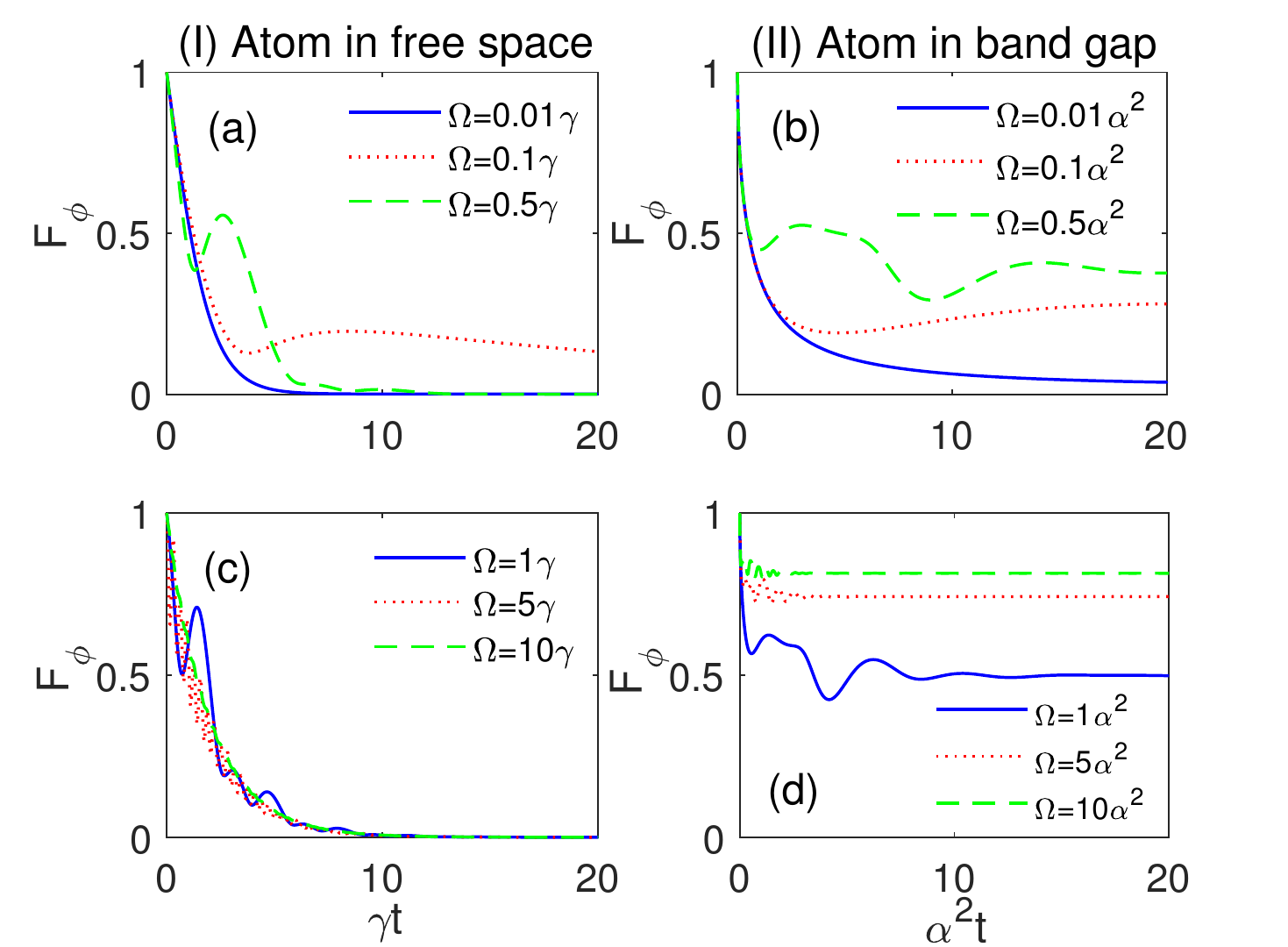}
\caption{Dynamical behavior of QFI $F_{\phi}$ for different Rabi frequencies of coupling laser field ($\Omega$), with $\theta=\pi/2$, $\phi=\pi/4$. Column (I) corresponds to the case in which atom is placed in free space and column (II) corresponds to the case when the atom is in a band-gap material with $\delta=0$.}
\label{fig6}
\end{figure}

\begin{figure}[t!]
\centering
\includegraphics[width=0.52\textwidth]{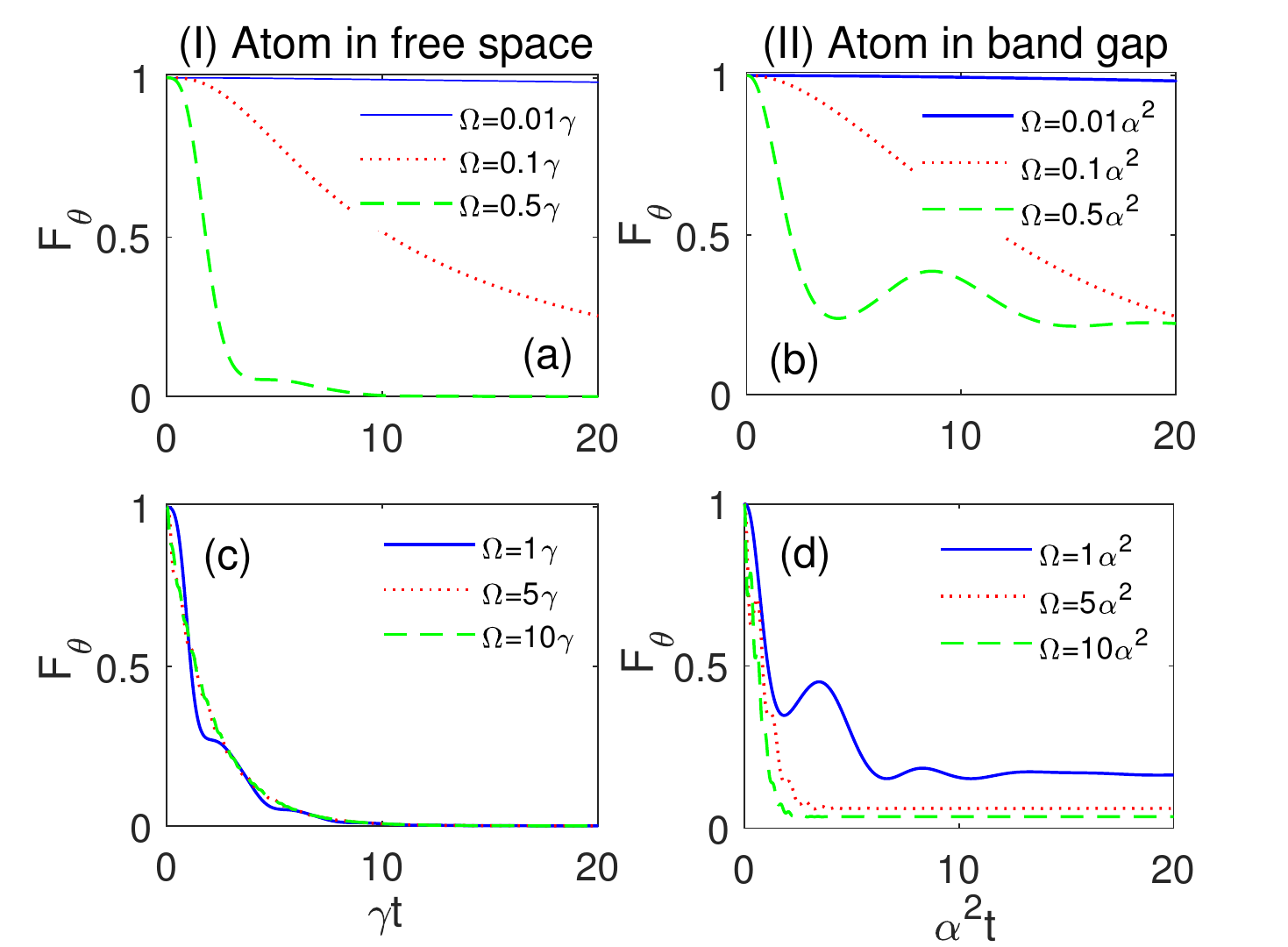}
\caption{Dynamical behavior of QFI $F_{\theta}$ for different Rabi frequencies of coupling laser field ($\Omega$), with $\theta=\pi/2$, $\phi=\pi/4$. Column (I) corresponds to the case in which atom is placed in free space and column (II) corresponds to the case in which atom in a band-gap material with $\delta=0$.}
\label{fig7}
\end{figure}

To meet the target, $F_{\phi}$ and $F_{\theta}$ are respectively depicted versus the scaled time for various values of $\Omega$ in Fig.~\ref{fig6} and Fig.~\ref{fig7}. The other parameters are chosen as $\theta=\pi/2$, $\phi=\pi/4$ and $\delta=0$. As can be observed in column (I) of Fig.~\ref{fig6}, when the atom is placed in the free space, $F_{\phi}$ quickly decays to zero for all the Rabi frequencies except for $\Omega=0.1\gamma$ for which the decay is slower. Larger $\Omega$ produces a nonmonotonic dynamics for the QFI.
For the atom placed in a photonic band gap (column (II) of Fig.~\ref{fig6}), $F_{\phi}$ decays and reaches a nonzero steady-state value. In this case, it is interesting to observe that increasing $\Omega$ not only causes the appearance of an oscillating evolution of $F_{\phi}$ but also accelerates the attainment of its stationary value. The larger $\Omega$, the larger the achieved steady-state value.
In Fig.~\ref{fig7} we report the dynamics of $F_{\theta}$, that responds in a completely different fashion to the increasing of $\Omega$ compared to $F_{\phi}$. For both media, one sees that larger values of $\Omega$ have detrimental effects on $F_{\theta}$. The most convenient condition for maintaining $F_{\theta}$ closer to its initial value is found for a weak coupling field, that is small values of $\Omega$. These radically different time behaviors of $F_{\theta}$ and $F_{\phi}$ are linked to the different role of the two angle parameters within the qutrit state: $\theta$ fixes the initial probability amplitudes of the state, while $\phi$ is a relative phase. The precision of the two different parameter estimations during the dynamics is differently influenced by the interaction of the atomic qutrit with the environment and the coupling field. Moreover, one expects that these time behaviors strongly depend on the values of the two initial parameters $\theta$ and $\phi$ to be measured.

\begin{figure}[t!]
\centering
\includegraphics[width=0.52\textwidth]{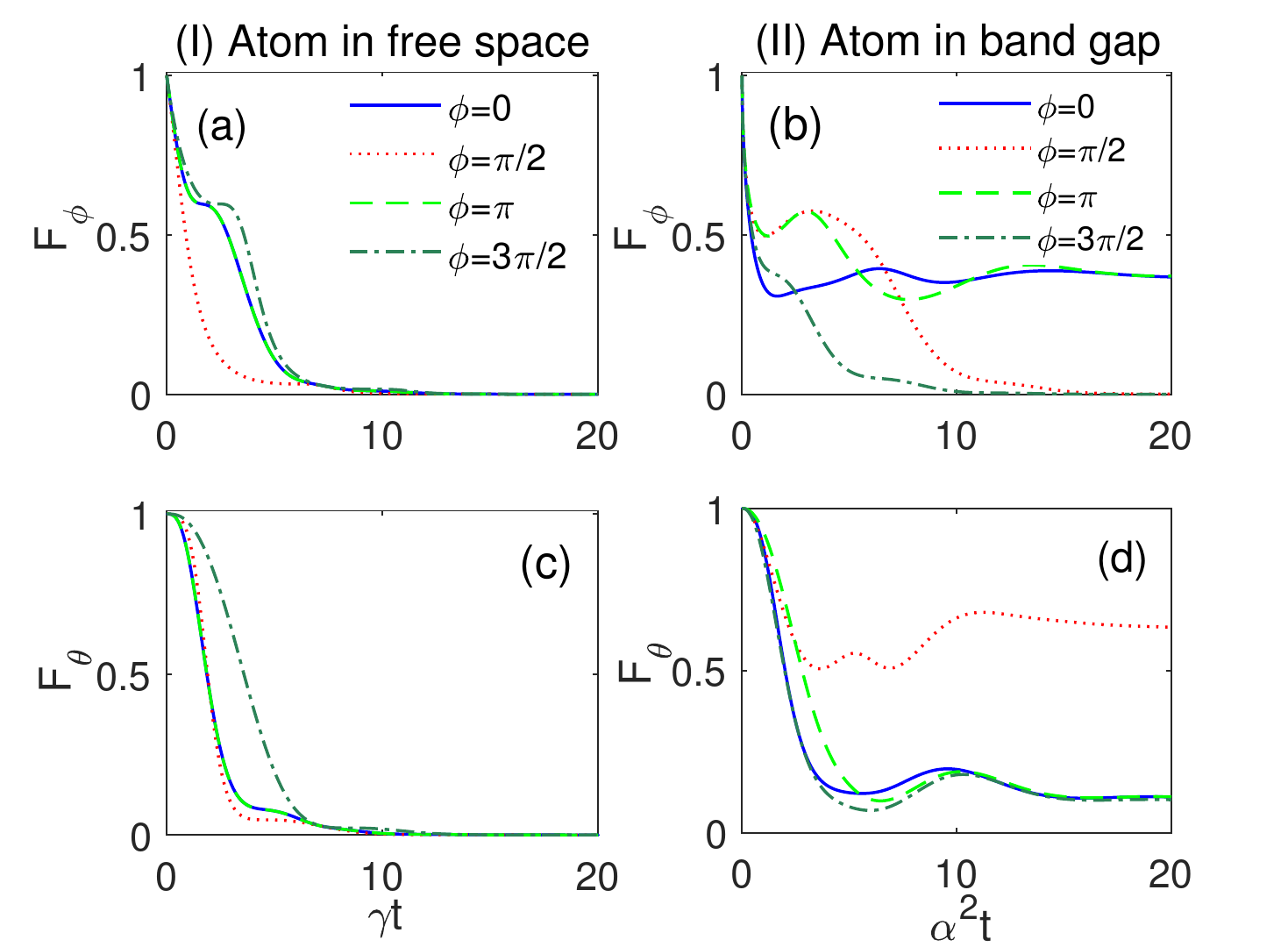}
\caption{Dynamical behavior of QFIs $F_{\phi}$ (upper row) and $F_{\theta}$ (lower row) for different values of $\phi$, with $\theta=\pi/2$. Column (I) corresponds to the atom in free space with $\Omega=0.5\gamma$. Column (II) corresponds to the atom in a band-gap material with $\Omega=0.5\alpha^2$, $\delta=0$.}
\label{fig8}
\end{figure}

To go deeper towards this relevant aspect in the quantum metrology scenario, we first study  the influence of the initial relative phase on the dynamics of $F_{\phi}$ and $F_{\theta}$, fixing $\theta=\pi/2$. In Fig.~\ref{fig8}, $F_{\phi}$ (panels (a), (b)) and $F_{\theta}$ (panels (c), (d)) are plotted as a function of the scaled time for various values of $\phi$. Column (I) regards the case of free space with $\Omega=0.5\gamma$ for the driving field, while column (II) the case of band-gap material with $\Omega=0.5 \alpha^2$. As expected for these values of $\Omega$, when the atom is in the free space both QFIs quickly tend to zero regardless of the initial phase value. On the other hand, when the atom is in the photonic band gap medium, $F_{\theta}$ tends to different nonzero steady state values (Fig.~\ref{fig8}(d)) for any $\phi$, which is not true in general for $F_{\phi}$ (Fig.~\ref{fig8}(b)). Some values of the initial relative phase can compromise the parameter estimation, causing $F_{\phi}$ to vanish. The initial phase value affects the evolution of $F_{\phi}$ for the specific case $\theta=\pi/2$. 

\begin{figure}[t!]
\centering
\includegraphics[width=0.4\textwidth]{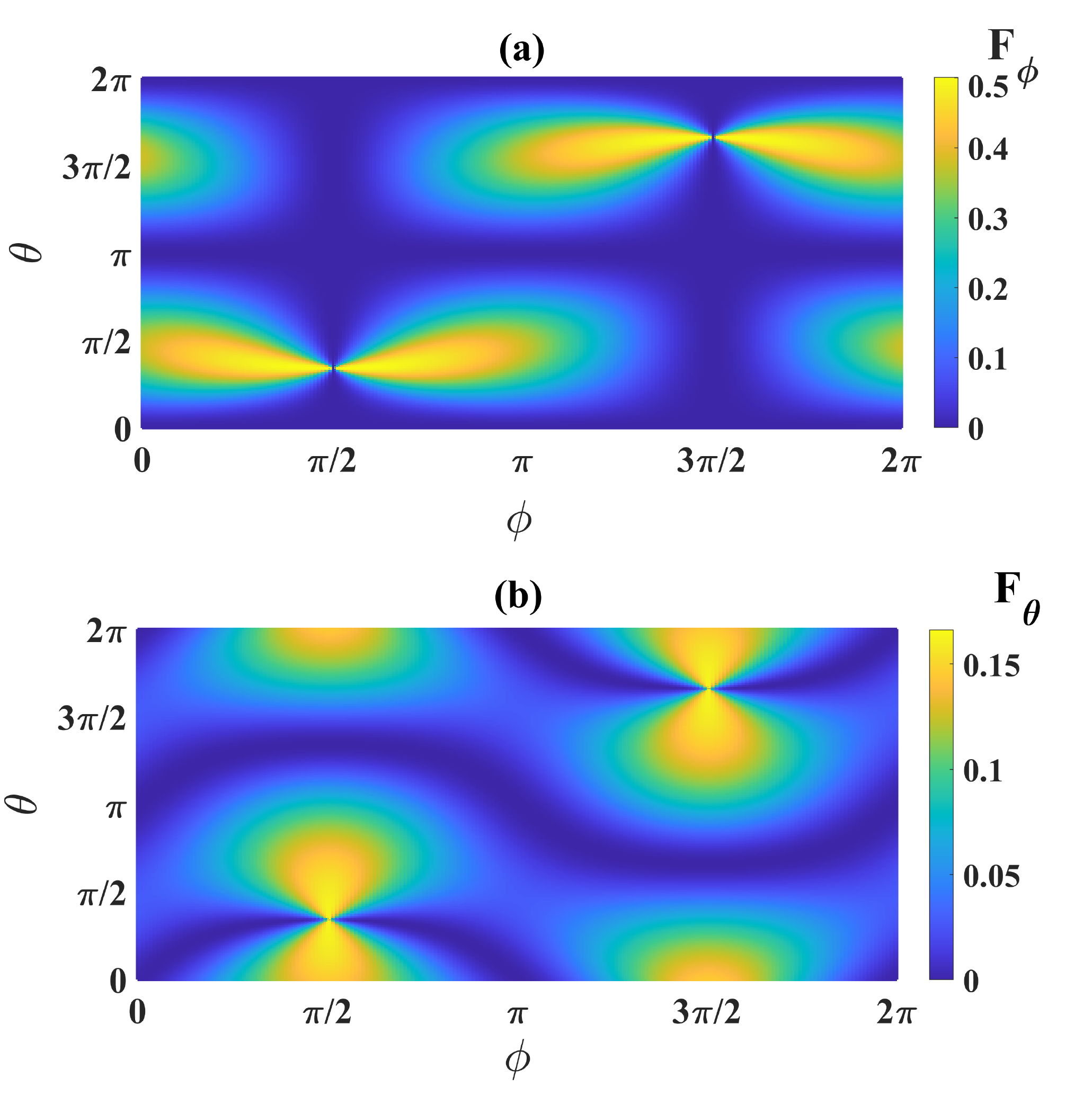}
\caption{Density Plot of the steady-state QFI (a) $F_{\phi}$ and (b) $F_{\theta}$ as functions of $\phi$ and $\theta$ when the atom is placed in a band-gap material. Values of parameters are: $ t \rightarrow \infty$, $\Omega=0.5\alpha^2$, $\delta=0$.}
\label{fig9}
\end{figure}

From the previous plots, we symmetrically expect that changing the value of $\theta$ for a given $\phi$ will also affect the evolution of the QFIs. To have a wider view of this feature, considering the advantageous case of photonic band gap reservoir, in Fig.~\ref{fig9} we report the steady-state values of $F_{\phi}$ and $F_{\theta}$ in a contour plot for a comprehensive range of initial state parameters $\phi$, $\theta$. The central role of the  latter in determining the steady-state values of $F_{\phi}$ and $F_{\theta}$ is evident. Appropriate choices of $\phi$ and $\theta$ can yield nonzero steady-state values for $F_{\phi}$ and $F_{\theta}$. For the chosen values of $\Omega$ and $\delta$ in these plots, we also notice that the steady-state value of $F_{\phi}$ is larger than $F_{\theta}$. In particular, for opportune combinations of initial angle parameters, one can reach the optimal condition for the stationary amount of $F_{\phi}=0.5$, enabling a sensitivity $\delta\phi_{min}=1/\sqrt{0.5}$ for the initial relative phase of the atomic state.

\subsection{Two-parameter Fisher information}

We now consider a two-parameter estimation problem and employ the QFIM approach to calculate the QCRB in the simultaneous estimation of both two parameters $\phi$ and $\theta$. 

\begin{figure}[t!]
\centering
\includegraphics[width=0.52\textwidth]{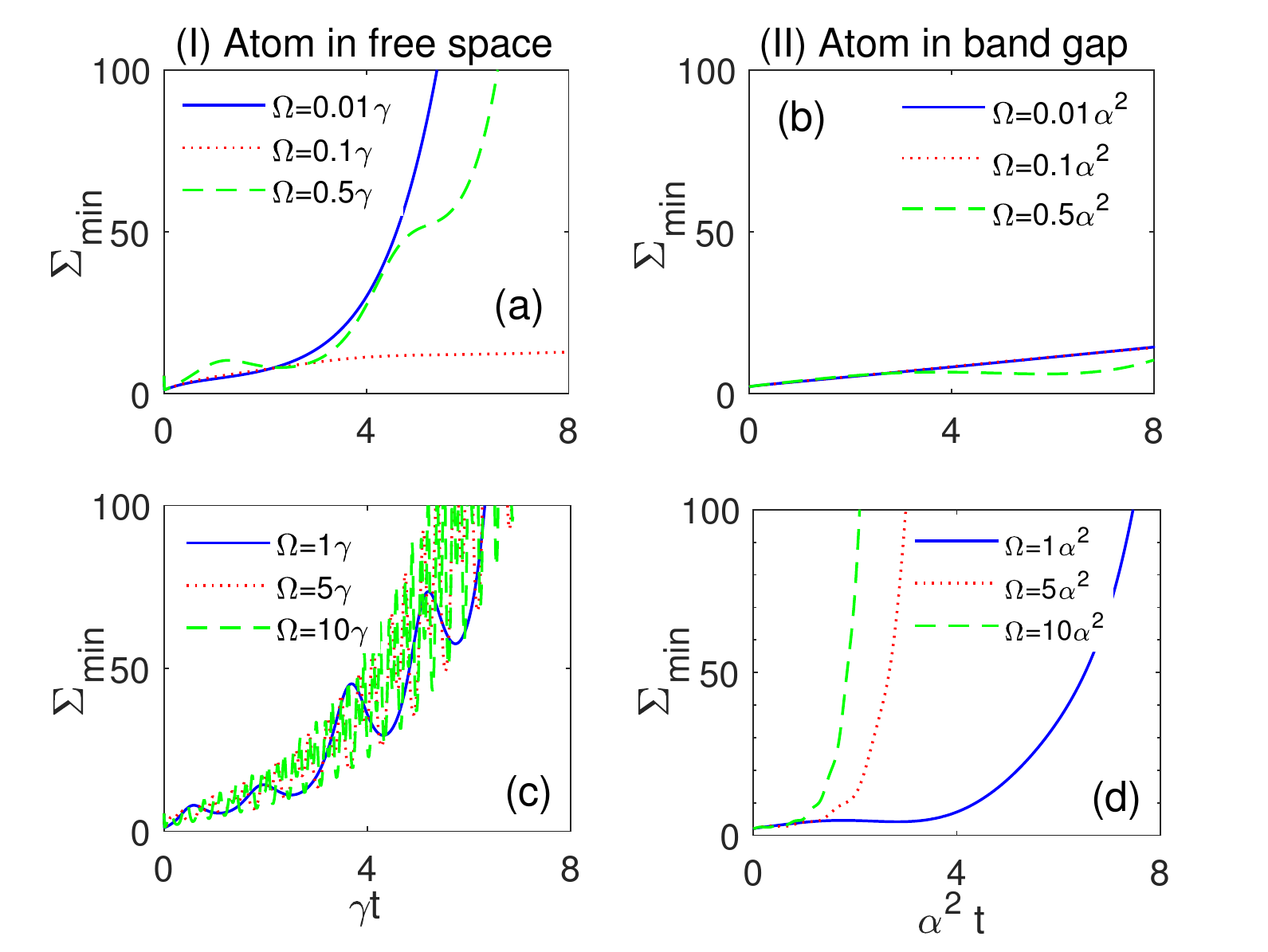}
\caption{Dynamical behavior of $\sum_{min}$ for different Rabi frequencies of coupling laser field ($\Omega$). Column (I) corresponds to the case in which atom is placed in free space and column (II) corresponds to the case in which the atom is in a band-gap material with $\delta=0$. Initial state parameters are fixed as $\theta=\pi/2$, $\phi=\pi/4$.}
\label{fig10}
\end{figure}

Along this route, we return to Eq.~\eqref{eq:34} and display $\sum_\mathrm{min}$ (optimal sensitivity) versus scaled time for different Rabi frequencies $\Omega$ of coupling laser field in Fig.~\ref{fig10}. In general, for both types of reservoir (free space and photonic crystal), $\sum_\mathrm{min}$ is going to increase as time goes by but with different rates depending on the Rabi frequency. For the case when the atomic qutrit is placed in free space (column I of Fig.~\ref{fig10}), it is clear that the best optimal two-parameter estimation can be acquired for $\Omega=0.1\gamma$; other values of $\Omega$ do not yield favorable results for the two-parameter estimation during the dynamics. In contrast, the band gap material as a reservoir (column II of Fig.~\ref{fig10}) enables a wider range of values of $\Omega$ (within about two orders of magnitude) which keep $\sum_\mathrm{min}$ small enough during the dynamics, with the most enduring results occurring for $\Omega\leq0.5\alpha^2$; for larger values of $\Omega$ this behavior is lost.

\begin{figure}[t!]
\centering
\includegraphics[width=0.5\textwidth]{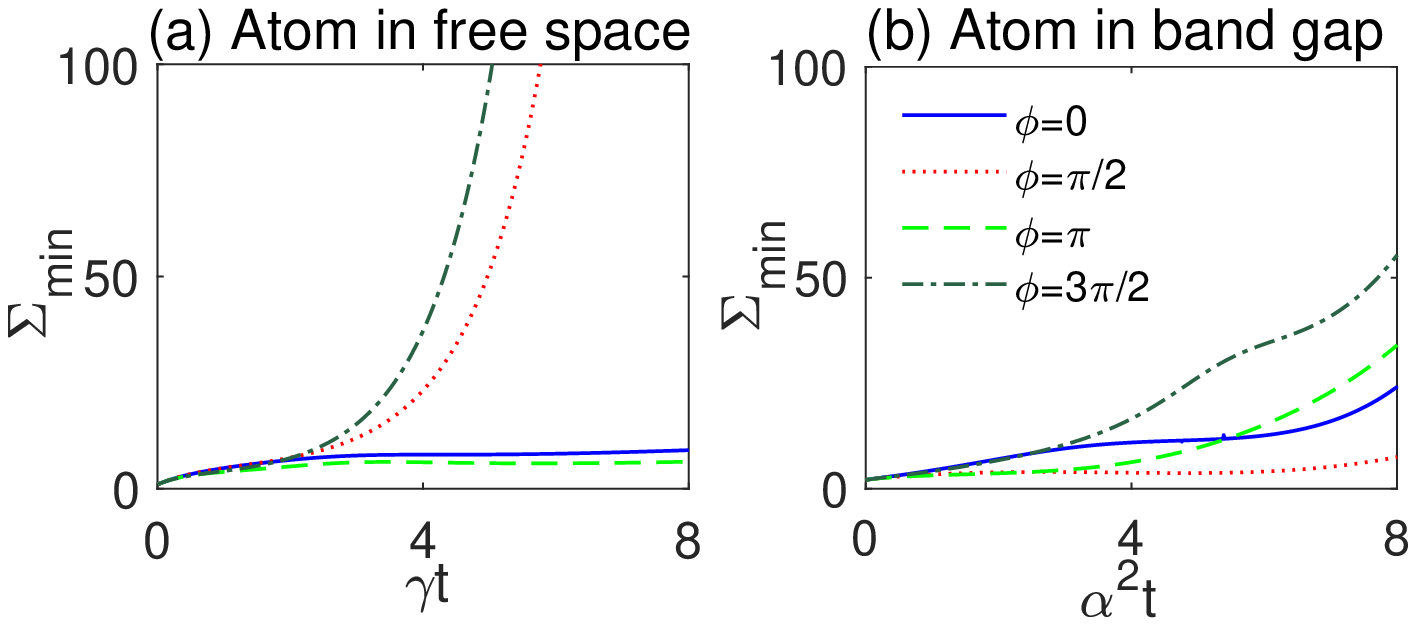}
\caption{Dynamical behavior of $\sum_{min}$ for different values of $\phi$, with $\theta=\pi/2$. (a) Atom in free space with $\Omega=0.1\gamma$; (b) atom in a band-gap material with $\Omega=0.5\alpha^2$, $\delta=0$.}
\label{fig11}
\end{figure}

Fig.~\ref{fig11} depicts the time evolution of $\sum_\mathrm{min}$ for various values of initial phase $\phi$ with $\theta=\pi/2$. We compare the two environmental conditions choosing the best case scenario for $\Omega$ as evinced from the above plots of Fig.~\ref{fig10}: $\Omega=0.1\gamma$ for free space, $\Omega=0.5\alpha^2$ for band gap material. In general, one sees that the initial phase value affects the results for both reservoirs. Notice that different values of $\phi$ give the most advantageous conditions for $\sum_\mathrm{min}$, e.g. $\phi=\pi$ for the free space and $\phi=\pi/2$ for the photonic crystal. A remarkable aspect can be however individuated: the case of photonic crystal keeps changes of $\sum_\mathrm{min}$ contained enough during the initial stages of the dynamics when $\phi$ varies, while changes of $\sum_\mathrm{min}$ can be abrupt for the case of free space. 

The main message of this analysis is the following: the interaction between the qutrit and the driving field must be in general suitably adjusted via $\Omega$ for achieving the best possible optimal sensitivity in the two-parameter measurement. 
Compared to the case of free space, a PBG material as structured environment interestingly allows a very large margin of error in setting the desired $\Omega$ and a minor dependence on variations of initial relative phase parameter $\phi$. This aspect may be especially relevant from an experimental viewpoint.

\section{Discussion on experimental feasibility}\label{sec:Disc}

Since the experimental validation of a theoretical study matters, we aim to find out the experimental feasibility of the proposed qutrit for both real and artificial atoms. For real case, the calcium atom exemplifies our model whose energy levels aptly map onto the considered qutrit \cite{Mortezapour1}. So that, level $4s^{2}\ {}^{1}S_{0}$ functions as ground state $\left| c \right\rangle$ level $4s3d\ {}^{1}D_{2}$ and $4s6p\ {}^{1}P_{1}$ which can be labeled as $\left| b \right\rangle$ and  $\left| a \right\rangle$ respectively correspond to metastable and excited states. In this case, the excited state $4s6p\ {}^{1}P_{1}$ can be coupled to level $4s3d\ {}^{1}D_{2}$ via a coupling field possessing a wavelength of 504 nm \cite{Mortezapour1}. Nobody could overlook the fact that fabrication of appropriate photonic crystal is equally important to embed the candidate qutrit into it. Although, the steady technical advancements have paved the way for fabrication of the photonic crystals having three-dimensional gap in both visible and infrared ranges \cite{PBGrev1,PBGrev2,PBGrev3}.

On the other hand, compared to real atoms, quantum dots (artificial atoms) have the merit to be readily coupled to photonic crystals. Solid-state nanostructures are attractive alternatives to atomic single photon emitters due to the fact that unlike real atoms, they do not require complex laser cooling and trapping techniques. In this regard, In recent years, significant progress has been achieved in the field of solid-state with three-level QD and photonic-crystal cavity \cite{QD1,QD2,QD3,QD4}. Meanwhile, a substantial number of three-level QDs have been put forward which received considerable attention \cite{QD5,QD6,QD7,QD8,QD9}.   

Albeit diverse three-level quantum dots have been fabricated so far, none of them perfectly conforms to our qutrit. The exceptional facet of our model that distinguishes it from the previous ones is its spontaneous decay rate from the excited state $\left| a \right\rangle$ to the metastable state $\left| b \right\rangle$ (i.e., $\gamma_{ab}$). So, the rate $\gamma_{ab}$ is negligible compared to the $\gamma_{ac}$ ($\gamma$) and the fabricated qutrits lack the required condition. Such controversial but appealing nature of the model poses a challenge to pondering possible ways to meet the constraint and find an appropriate quantum dot for implementation of the qutrit. 

One of the neat possible solutions is to consider a quantum dot qubit whose excited state is connected to a level of another quantum dot via tunneling effect which leads to the formation of a so-called quantum dot molecule \cite{Tunel1,Tunel2,Tunel3,Tunel4,Tunel5}. Such quantum dot molecule constitutes a way-out to the problem provided that the decay rate of the coupled level (second quantum dot) is small enough. The experimental realization of this particular qutrit does not seem far-fetched and is still open to debate.

\section{Conclusions}\label{sec:Conc}

A comparative study of a classically driven three-level atomic system (qutrit) placed in either free space (Markovian environment) or a photonic band gap crystal (structured environment) has been carried out. The aim of the study is to assess the impact of a classical driving field on the time evolution of relevant quantum properties, such as quantum coherence, non-Markovianity, and quantum Fisher information (QFI), which are encoded initially in the qutrit state. The study provides quantitative evidence that PBG materials can be employed as an effective environment in which all the achieved behaviors concerning non-Markovianity increase, coherence protection and quantum parameter estimation supersede those in the case of free space. 

The initial state of the atomic qutrit has been defined by two angle parameters, namely $\theta$ (ruling the probability amplitudes) and $\phi$ (relative phase between basis states). We have found that the initial relative phase has a substantial impact on the steady-state value of coherence. In particular, half of the initial coherence can be maintained by well-adjusting $\phi$. 

The angle parameters of the initial state can be unknown and therefore subject to precision measurements for their estimation. Within this quantum metrology context, we have investigated the QFI matrix both for single-parameter estimation, namely $F_{\phi}$ and $F_{\theta}$ individually, and for the simultaneous two-parameter estimation. Interestingly, it has been observed that $F_{\phi}$ and $F_{\theta}$ exhibit opposite behaviors versus variations of the intensity of the classical field. In particular, increasing the Rabi frequency of the classical driving field leads to increasing (decreasing) the steady-state value of $F_{\phi}$ ($F_{\theta}$). In other words, a larger Rabi frequency can enrich the sensitivity of phase estimation ($\phi$) in the steady state; instead, a more precise estimation of $\theta$ occurs as a result of decreasing the Rabi frequency. Moreover, we have reported that larger values of the Rabi frequency always lead to deterioration of simultaneous optimal two-parameter estimation. As a general trait, we have found that the values of relative phase $\phi$ affect the precision of both single and two-parameter estimation outcomes.

We have finally discussed the experimental context where the findings of this work can be reproduced, showing that systems made of controlled quantum dots in photonic crystals appear to be the most promising platforms. Ultimately, we have demonstrated that the cooperative utilization of a PBG material as medium and a classical driving field as part of the system enhances the quantum features of the qutrit during the dynamics.  In conclusion, our results provide useful insights towards the development of techniques for preserving the quantum properties of qutrit-based compounds in a quantum information scenario.

\section*{ACKNOWLEDGEMENTS}
A.M. acknowledges the support of the University of Guilan. R.L.F. acknowledges support from ``Sistema di Incentivazione, Sostegno e Premialit\`a della Ricerca Dipartimentale'', Dipartimento di Ingegneria, Universit\`a di Palermo.


%

 \end{document}